\documentclass[peerreview,12pt]{IEEEtran}
\usepackage{amsfonts,graphicx,enumerate,amsmath,amssymb}
\usepackage{mathtools,lipsum,placeins,fixltx2e}
\usepackage[inline]{enumitem}

\usepackage{cuted}
\newcommand{\EE}{\mathrm{E}}

\usepackage{setspace}
\begin{document}

\title{Higher-Degree Stochastic Integration Filtering}
\author{Syed~Safwan~Khalid,~Naveed~Ur~Rehman,~and~Shafayat~Abrar
\thanks{SSK and NUR are affiliated with COMSATS Institute of Information Technology, Dept. of Electrical
Engineering, Islamabad, 44000, Pakistan (Email: \{safwan\_khalid,naveed.rehman\}@comsats.edu.pk, shafayat1972@yahoo.com).}}

\maketitle


\begin{abstract}
We obtain a class of higher-degree stochastic integration filters (SIF) for nonlinear filtering applications. SIF are based on stochastic spherical-radial integration rules that achieve asymptotically exact evaluations of Gaussian weighted multivariate integrals found in nonlinear Bayesian filtering. The superiority of the proposed scheme is demonstrated by comparing the performance of the proposed fifth-degree SIF against a number of existing stochastic, quasi-stochastic and cubature (Kalman) filters. The proposed filter is demonstrated to outperform existing filters in all cases. 	
\end{abstract}

\section{Introduction}

Bayesian filtering provides a theoretical framework for recursive estimation of unknown dynamic state vectors in linear/nonlinear  filtering applications. In  Bayesian paradigm, the posterior probability of the state vector given the noisy observations is recursively updated at each instant. However, in general, the evaluation of the posterior probability is analytically intractable, and hence only approximate solutions are available \cite{haug2012bayesian}. The approximation methods are generally divided broadly into two categories, i.e., the global and the local methods \cite{arasaratnam2009cubature}. In the global approach, no assumption is made regarding the distribution of the posterior density and it is approximated using methods such as particle filtering \cite{gordon1993novel}, Gaussian mixtures \cite{alspach1972nonlinear} and point-mass filtering \cite{vsimandl2006advanced} etc. The filters in this category -- despite being fairly accurate -- are known to suffer from enormous computational load. 

On the other hand, methods based on the local approach are computationally less demanding. These methods rely on the assumption that the required posterior probability is Gaussian; consequently, the task of filtering is simplified to the recursive updates of the first- and the second-order moments only. The moment update relations essentially require solution of Gaussian weighted integrals of nonlinear functions. One possible approach is to use approximations such as Taylor series \cite{schmidt1981kalman}, Stirling's interpolation \cite{vsimandl2009derivative}, Fourier-Hermite series \cite{sarmavuori2012fourier} etc., that would make Gaussian integral tractable. Another possibility is to apply numerical integration methods to evaluate Gaussian weighted integrals \cite{wu2006numerical} thus giving rise to a large class of sigma-point Kalman filters e.g., the cubature Kalman filter (CKF) \cite{arasaratnam2009cubature}, the unscented Kalman filter \cite{uhlmann2000new}, the Gauss-Hermite quadrature filter  \cite{ito2000gaussian} etc. Using Monte-Carlo based stochastic numerical integration rules is another possibility resulting in Monte-Carlo Kalman filter (MCKF) \cite{song2000monte}.  Recently, in \cite{dunik2013stochastic} a stochastic integration filter (SIF) based on the third-degree stochastic spherical-radial rule was presented that provided asymptotically exact integral evaluations with faster convergence as compared to MCKF. The SIF can be considered as a stochastic counterpart of third-degree CKF. The inadequacy of third-degree integration rules in problems involving high nonlinearities and large uncertainties has been noted in the works of Jia et al. \cite{jia2011sparse, jia2012sparse}. Consequently, in the past few years, many researchers have focused their efforts on the development of higher-degree cubature Kalman filters \cite{jia2013high, wang2014spherical, zhang2014seventh}. The motivation behind our work is to discuss the development and performance of higher-degree stochastic counterparts of these cubature filters. We first describe stochastic integration rules for an arbitrary degree, and then proceed to develop a fifth-degree SIF.

This paper is organized as follows: Section \ref{sec: Bayesian Framework} describes Bayesian filtering briefly. Section \ref{sec: SIF} presents stochastic spherical-radial (integration) rule of a generic degree. Section \ref{SectionSIF} proposes a fifth-degree stochastic integration rule for Bayesian filtering.
Section \ref{sec: Simulations} presents simulation results, and Section \ref{sec: Conclusions} draws conclusions.

\section {Bayesian Filtering Framework}
\label{sec: Bayesian Framework}

Consider a representative nonlinear system:
\begin{subequations}
\label{Eq: System Models}
\begin{alignat}{1}
\boldsymbol{x}_k &= f(\boldsymbol{x}_{k-1}) + \boldsymbol{w}_k, \\
\boldsymbol{y}_k &= h(\boldsymbol{x}_k) + \boldsymbol{v}_k,
\end{alignat}
\end{subequations}
where \(\boldsymbol{x}_k \in \mathbb{R}^n\) and \(\boldsymbol{y}_k \in \mathbb{R}^m\) are state and observation vectors, respectively. The system model \(f(\cdot)\) and the observation model \(h(\cdot)\) are nonlinear functions. The noise processes \(\boldsymbol{w}_k\) and \(\boldsymbol{v}_k\) represent the uncertainties in the models and are zero mean Gaussian random processes, i.e., \(\boldsymbol{w}_k \sim \mathcal{N}(\boldsymbol{0};Q_k)\) and \(\boldsymbol{v}_k \sim \mathcal{N}(\boldsymbol{0};R_k)\). Let \(\boldsymbol{Y}_k = \{\boldsymbol{y}_0,\boldsymbol{y}_1,\cdots,\boldsymbol{y}_k\}\) be the set of all available observations at
\(k\)th instant. The aim of filtering process is to provide an estimate of the state vector given \(\boldsymbol{Y}_k\). We know that the
optimal estimate in terms of minimum mean square error (MSE) is given by \(\hat{\boldsymbol{x}}_{k|k} = \EE[\boldsymbol{x}_k|\boldsymbol{Y}_k]\), i.e., \(\hat{\boldsymbol{x}}_{k|k} = \int\boldsymbol{x}_kp(\boldsymbol{x}_k|\boldsymbol{Y}_k)\textrm{d}\boldsymbol{x}_k\). Using Bayes theorem, we get \(p(\boldsymbol{x}_k|\boldsymbol{Y}_k) = \frac{1}{c}{p(\boldsymbol{y}_k|\boldsymbol{x}_k)p(\boldsymbol{x}_k|\boldsymbol{Y}_{k-1})}\), where \(p(\boldsymbol{x}_k|\boldsymbol{Y}_{k-1}) = \int p(\boldsymbol{x}_k|\boldsymbol{x}_{k-1})p(\boldsymbol{x}_{k-1}|\boldsymbol{Y}_{k-1})\textrm{d}\boldsymbol{x}_{k-1}\), and \(c:={p(\boldsymbol{y}_k|\boldsymbol{Y}_{k-1})}\). Hence we have a recursive relation to evaluate \(p(\boldsymbol{x}_k|\boldsymbol{Y}_k)\) and consequently \(\hat{\boldsymbol{x}}_{k|k}\). Assuming that \(p(\boldsymbol{x}_k|\boldsymbol{Y}_k) = \mathcal{N}_{\boldsymbol{x}_k}(\hat{\boldsymbol{x}}_{k|k},P^{xx}_{k|k})\) and \(p(\boldsymbol{x}_k|\boldsymbol{Y}_{k-1}) = \mathcal{N}_{\boldsymbol{x}_k}(\hat{\boldsymbol{x}}_{k|k-1},P^{xx}_{k|k-1})\), the optimal estimate \(\hat{\boldsymbol{x}}_{k|k}\) admits a solution \cite{haug2012bayesian}, see (\ref{Eq: x hat k given k})-(\ref{Eq: P xx k given k}).
\begin{table}
\centering
\begin{minipage}{0.495\textwidth}
\hrule\vspace{-2mm}
\begin{align}\nonumber & \hspace{-7mm}{\normalsize\underline{\textrm{State prediction step:}}}\\
\label{Eq: x hat k given k}
\hat{\boldsymbol{x}}_{k|k-1} \!&= \EE \big[\boldsymbol{x}_k|\boldsymbol{Y}_{\!\!k-1}\big]\! = \!\!\int\!\!\! f(\boldsymbol{x}_{k-1}) \mathcal{N}_{\boldsymbol{x}_{k-1}}(\hat{\boldsymbol{x}}_{k-1|k-1},P^{xx}_{k-1|k-1})\textrm{d}\boldsymbol{x}_{k-1}\\
\nonumber P^{xx}_{k|k-1} & = \EE\big[\big(\boldsymbol{x}_k - \hat{\boldsymbol{x}}_{k|k-1}\big)\big(\boldsymbol{x}_k - \hat{\boldsymbol{x}}_{k|k-1}\big)^T|\boldsymbol{Y}_{k-1}\big] \\
\nonumber &= Q_k-\hat{\boldsymbol{x}}_{k|k-1}\hat{\boldsymbol{x}}_{k|k-1}^T + \\ & \label{eq: P k given k-1} \int f(\boldsymbol{x}_{k-1})f(\boldsymbol{x}_{k-1})^T\mathcal{N}_{\boldsymbol{x}_{k-1}}(\hat{\boldsymbol{x}}_{k-1|k-1},P^{xx}_{k-1|k-1})\textrm{d}\boldsymbol{x}_{k-1}, \\ \nonumber &\hspace{-7mm} {\normalsize\underline{\textrm{Observation prediction step:}}}\\
 \label{eq: y hat k given k-1}\hat{\boldsymbol{y}}_{k|k-1} & = \EE\big[\boldsymbol{y}_k|\boldsymbol{x}_k,\boldsymbol{Y}_{k-1}\big] = \int h(\boldsymbol{x}_k) \mathcal{N}_{\boldsymbol{x}_{k}}(\hat{\boldsymbol{x}}_{k|k-1},P^{xx}_{k|k-1})
\textrm{d}\boldsymbol{x}_{k},\\
\nonumber P^{xy}_{k|k-1} &= \EE\big[\big(\boldsymbol{x}_k - \hat{\boldsymbol{x}}_{k|k-1}\big)\big(\boldsymbol{y}_k - \hat{\boldsymbol{y}}_{k|k-1}\big)^T\big|\boldsymbol{x}_k,\boldsymbol{Y}_{k-1}\big] \\
 & \label{eq: P xy k given k-1} = \int \boldsymbol{x}_k h(\boldsymbol{x}_k)^T \mathcal{N}_{\boldsymbol{x}_{k}}(\hat{\boldsymbol{x}}_{k|k-1},P^{xx}_{k|k-1})
\textrm{d}\boldsymbol{x}_{k}  - \hat{\boldsymbol{x}}_{k|k-1}\hat{\boldsymbol{y}}_{k|k-1}^T,\\
\nonumber P^{yy}_{k|k-1} &= \EE\big[\big(\boldsymbol{y}_k - \hat{\boldsymbol{y}}_{k|k-1}\big)\big(\boldsymbol{y}_k - \hat{\boldsymbol{y}}_{k|k-1}\big)^T|\boldsymbol{x}_k,\boldsymbol{Y}_{k-1}\big] \\
 &\hspace{-8mm} \label{Eq: P yy k given k minus 1} = \int h(\boldsymbol{x}_k)h(\boldsymbol{x}_k)^T \mathcal{N}_{\boldsymbol{x}_{k}}(\hat{\boldsymbol{x}}_{k|k-1},P^{xx}_{k|k-1})\textrm{d}\boldsymbol{x}_{k} - \hat{\boldsymbol{y}}_{k|k-1}\hat{\boldsymbol{y}}_{k|k-1}^T +R_k\\
 \nonumber &\hspace{-7mm} {\normalsize\underline{\textrm{Bayesian filter correction step:}}}\\
 \label{Eq: x k given k}\hat{\boldsymbol{x}}_{k|k} &= \hat{\boldsymbol{x}}_{k|k-1} + P^{xy}_{k|k-1}[P^{yy}_{k|k-1}]^{-1}\big(\boldsymbol{y}_k - \hat{\boldsymbol{y}}_{k|k-1}\big),\\
\label{Eq: P xx k given k} P^{xx}_{k|k} & = P^{xx}_{k|k-1} - P^{xy}_{k|k-1}[P^{yy}_{k|k-1}]^{-1}\big[P^{xy}_{k|k-1}\big]^T.
\end{align}
\hrule
\end{minipage}
\end{table}

Note that the Bayesian filtering process essentially breaks down to the evaluation of Gaussian weighted integrals of the form \(I(s) = \int s(\boldsymbol{x})\mathcal{N}_{\boldsymbol{x}}(\hat{\boldsymbol{x}},P)\textrm{d}\boldsymbol{x}\). The integral \(I(s)\), in general, does not admit a closed-form solution, and thus, numerical integration is employed \cite{arasaratnam2009cubature,wu2006numerical,ito2000gaussian,dunik2013stochastic}.
\section{Stochastic Integration Method}
\label{sec: SIF}
Here, we describe stochastic integration method of arbitrary accuracy to approximate the Gaussian weighted integral \(I(s)\), and consequently develop a fifth-degree stochastic integration (Bayesian) filter.

We introduce a transformation \(\boldsymbol{x} = \hat{\boldsymbol{x}} + \sqrt{P}\boldsymbol{c}\), where \(P = \sqrt{P}\sqrt{P}^T\) \cite{haug2012bayesian}; accordingly, the Gaussian weighted integral is written as \(\int {s}(\hat{\boldsymbol{x}} + \sqrt{P}\boldsymbol{c})\mathcal{N}_{\boldsymbol{c}}(\boldsymbol{0},I)\textrm{d}\boldsymbol{c}=\int {g}(\boldsymbol{c})\mathcal{N}_{\boldsymbol{c}}(\boldsymbol{0},I)\textrm{d}\boldsymbol{c}=:I(g)\), where \(g(\boldsymbol{c}):=s(\hat{\boldsymbol{x}} + \sqrt{P}\boldsymbol{c})\). Secondly, we introduce a change of variable to convert the integral into the radial-spherical coordinate system, i.e., we let \(\boldsymbol{c} = r\boldsymbol{z}\), with \(\boldsymbol{z}\boldsymbol{z}^T = 1\), \(w(||\boldsymbol{c}||) := (2\pi)^{-\frac{{n}}{2}} \exp(-\frac{1}{2}\boldsymbol{c}^T\boldsymbol{c})\), and \(r^2 = \boldsymbol{c}^T\boldsymbol{c}\) \cite{genz1998stochastic},
\begin{equation}
\label{Eq: I of g}
I(g) = \int_{\boldsymbol{z}^T\boldsymbol{z}} \int_{0}^{\infty}w(r)r^{n-1}g(r\boldsymbol{z})\textrm{d}r\textrm{d}\boldsymbol{z},
\end{equation}
where \(w(r) = (2\pi)^{-\frac{n}{2}}\exp(-\frac{1}{2}{r^2})\). We approximate the radial integral using a stochastic radial rule of the form
\begin{subequations}
\begin{alignat}{1}
\label{Eq: Radial Integral}
I_r(g) &= \int_{0}^{\infty} w(r)r^{(n-1)}g(r)\textrm{d}r\\
\label{Eq: Radial Rule}
 &\approx \sum_{i=0}^{N_r} \varpi_{r,i} \!\left[\frac{g(\rho_i) + g(-\rho_i)}{2}\right]
\end{alignat}
\end{subequations}
where weights \(\{\varpi_{r,i}\}\) with a set of random points \(\{\rho_i\}\) are selected such that (\ref{Eq: Radial Rule}) becomes a \(d\)th-degree integration rule for (\ref{Eq: Radial Integral}).
Similarly, we have a spherical rule
\begin{equation}
\label{Eq: Spherical Rule}
I_{\boldsymbol{z}}(g) = \int_{\boldsymbol{z}^T\boldsymbol{z}} g(\boldsymbol{z})\textrm{d}\boldsymbol{z} \approx \sum_{j=0}^{N_s}\varpi_{s,j}g(\mathcal{Q}\boldsymbol{z}_j).
\end{equation}
Combining (\ref{Eq: Radial Rule}) and (\ref{Eq: Spherical Rule}), a product stochastic spherical-radial rule is defined to approximate \(I(g)\), i.e.,
\begin{equation}
\label{Eq: Spherical Radial Rule}
I(g) \approx \sum_{j=0}^{N_s}\varpi_{s,j}\sum_{i=0}^{N_r}\varpi_{r,i}\left[\frac{g(\rho_i\mathcal{Q}\boldsymbol{z}_j) + g(-\rho_i\mathcal{Q}\boldsymbol{z}_j)}{2}\right].
\end{equation}
where \(\{\varpi_{s,j}\}\) are weights, and \(\mathcal{Q}\) is an orthogonal matrix.

\textit{Remark 1}: The spherical-radial rule described above is a \(d\)th-degree rule  if  \begin {enumerate*}  \item it is exact for a \(g(\boldsymbol{x})\) that can be described by a linear combination of monomials up to degree $d$, \item It is not exact for at least one monomial of degree $d + 1$ \end{enumerate*}. Moreover, if the radial rule in (\ref{Eq: Radial Rule}) and the spherical rule in (\ref{Eq: Spherical Rule}) are both \(d\)th-degree, then the resulting spherical-radial rule in (\ref{Eq: Spherical Radial Rule}) is \(d\)th-degree as well \cite{jia2013high}.

\subsection{Stochastic Radial Rule}
To realize the radial rule (\ref{Eq: Radial Rule}), we have a proposition:
\smallskip

\textit{Proposition 1 \cite{genz1998stochastic}}: \textit{If weights \(\varpi_{r,i}\) in (\ref{Eq: Radial Rule}) are defined by
\begin{small}\begin{equation}
\label{Eq: weights general formula}
\varpi_{r,i} = I_r\left(\prod_{k=0,k \neq i}^{N_r}\frac{r^2 - \rho_k^2}{\rho_i^2 - \rho_k^2}\right),
\end{equation}\end{small}
where \(\rho_0 = 0\) and \(\rho_i\) is chosen from a distribution proportional to \(p(\rho_1,\rho_2,\cdots,\rho_{N_r}) = \prod_{i=1}^{N_r}\rho_i^{n+1}w(\rho_i)\prod_{k=1}^{i-1}(\rho_i - \rho_k)^2(\rho_i+\rho_k)\), then (\ref{Eq: Radial Rule}) is an unbiased degree \(2N_r+1\) integration rule for \(I_r(g)\).}

\smallskip
\textit{Remark 2}: Note that, it is not straightforward to sample the distribution \(p(\rho_1,\rho_2,\cdots,\rho_{N_r})\) for an arbitrary \(N_r\). For \(N_r = 1\), the required probability is \(p(\rho_1) \propto (\rho_1)^{n+1}\exp(-\rho_1^2/2)\), i.e., a chi-distribution with \(n+2\) degrees of freedom. For \(N_r = 2\), \(p(\rho_1,\rho_2) \propto (\rho_1\rho_2)^{n+1}\exp(-\frac{1}{2}{(\rho_1^2+\rho_2^2)})(\rho_2 - \rho_1)^2(\rho_2 + \rho_1)\). The probability \(p(\rho_1,\rho_2)\) is not a standard distribution; however, if we choose some \(\eta_1\) from chi-distribution with \(2n+7\) degrees of freedom, and some \(\eta_2\) from beta-distribution with \(\alpha = n+2\) and \(\beta = \frac{3}{2}\), then \(\rho_1 = \eta_1\sin(\frac{1}{2}{\sin^{-1}(\eta_2)})\) and \(\rho_2 = \eta_1\cos(\frac{1}{2}\sin^{-1}(\eta_2))\) will be distributed proportional to \(p(\rho_1,\rho_2)\) \cite{genz1998stochastic}. For \(N_r \geq 3\), the resulting joint distributions are either not standard or not easily factored into standard forms, and hence methods like Monte-Carlo sampling, such as rejection sampling \cite{liu2008monte}, may be employed.

\subsection{Stochastic Spherical Rule}

A large variety of deterministic integration rules are available in literature to approximate the spherical integral \(I_{\boldsymbol{z}}(g)\). For instance, \cite{jia2013high} describes a method to develop spherical rules of arbitrary degrees based on the work of Genz \cite{genz2003fully}. More efficient fifth- and seventh-degree rules can be found in \cite{lu2004higher} and \cite{stoyanova1997cubature}, respectively. Here, however, we are interested in converting a given deterministic rule into a stochastic one. To do so, we exploit the following proposition:

\smallskip
\textit{Proposition 2 \cite{genz1998stochastic}}: \textit{Let \(S(g) = \sum_{j=0}^{N_s}\varpi_{s,j}g(\boldsymbol{z}_j)\) be an integration rule of degree \(d\) for the integral \(I_{\boldsymbol{z}}(g)\). If \(\mathcal{Q}\) is a uniformly chosen \(n \times n\) orthogonal matrix, then \(S_{\mathcal{Q}}(g) = \sum_{j=0}^{N_s}\varpi_{s,j}g(\mathcal{Q}\boldsymbol{z}_j)\) is also an unbiased integration rule of degree \(d\) for \(I_{\boldsymbol{z}}(g)\).}

\smallskip
\textit{Remark 3}: We can develop a stochastic spherical rule of an arbitrary degree using \textit{Proposition 2} and any of the various rules available in the literature \cite{genz2003fully, lu2004higher, stoyanova1997cubature}. The standard method for generating \(\mathcal{Q}\) is to set it equal to the \(Q\) matrix of the \(QR\)-factorization of an \(n \times n\) random matrix \(X\), where each entry of \(X\) is independent and distributed in \(\mathcal{N}(0,1)\). More efficient methods can be found in \cite{genz1998methods}.

\subsection{Fifth-degree Stochastic Spherical Radial Rule}
\label{subsec: Fifth-degree rule}
To develop a fifth-degree stochastic radial rule (\(N_r =2\)), we note from \textit{Proposition 1} that the corresponding weights \(\varpi_{r,0}\), \(\varpi_{r,1}\) and \(\varpi_{r,2}\) are evaluated as follows:
\begin{small}
\begin{subequations}
\label{Eq: Weights formulae}
\begin{alignat}{1}
\varpi_{r,0} &= I_r\left(\frac{(r^2 - \rho_1^2)(r^2 - \rho_2^2)}{\rho_1^2\rho_2^2}\right) = T\bigg[1 - \frac{n(\rho_1^2 + \rho_2^2 - (n+2))}{\rho_1^2\rho_2^2}\bigg]\\
\varpi_{r,1} &= I_r\left(\frac{r^2(r^2 - \rho_2^2)}{\rho_1^2(\rho_1^2 - \rho_2^2)}\right) = T\frac{n(n+2-\rho^2_2)}{\rho_1^2(\rho_1^2 - \rho_2^2)}\\
\varpi_{r,2} &= I_r\left(\frac{r^2(r^2 - \rho_1^2)}{\rho_2^2(\rho_2^2 - \rho_1^2)}\right) = T\frac{n(n+2-\rho^2_1)}{\rho_2^2(\rho_2^2 - \rho_1^2)}
\end{alignat}
\end{subequations}
\end{small}
where \(T=\pi^{-{n}/{2}}\Gamma(n/2)\). The method for generating \(\rho_1,\rho_2\) has been discussed in \textit{Remark 2}.

\smallskip
For the fifth-degree stochastic spherical rule, we first employ the deterministic spherical-simplex method  \cite{wang2014spherical,lu2004higher} and then make use of \textit{Proposition 2} to convert it into a stochastic rule. The spherical-simplex rule is given as:
\begin{small}
\begin{equation}
\label{Eq: Spherical Simplex}
\begin{aligned}
I_{\boldsymbol{z}}(g)  \approx \frac{2\varpi_{s,1}}{T}\sum_{j=1}^{n+1}\big[g(\boldsymbol{a}_j) + g(-\boldsymbol{a}_j)\big]   + \frac{2\varpi_{s,2}}{T}\sum_{j=1}^{n(n+1)/2}\big[g(\boldsymbol{b}_j) + g(-\boldsymbol{b}_j)\big],
\end{aligned}
\end{equation}
\end{small}
where \(2/T\) is the surface area of unit sphere, the weights are given as \(\varpi_{s,1} = \frac{(7-n)n}{2(n+1)^2(n+2)}\) and \(\varpi_{s,2} = \frac{2(n-1)^2}{n(n+1)^2(n+2)}\). The vector points \(\boldsymbol{a}_j = [a_{j,1},a_{j,1},\cdots,a_{j,n}]^T\) are the vertices of an $n$-simplex and are given as
\begin{equation}
\label{Eq: a j}
a_{j,k} =\left\{\begin{array}{lr}
-\sqrt{\frac{n+1}{n(n-k+2)(n-k+1)}}, & k<j\\
+\sqrt{\frac{(n+1)(n-j+1)}{n(n-j+2)}}, & k=j \\
0, & k>j
\end{array}\right.
\end{equation}
Whereas, \(\boldsymbol{b}_j\) are the midpoints of \(\boldsymbol{a}_j\) projected onto the spherical surface, i.e.,
\(\boldsymbol{b}_j = \sqrt{{n}/({2(n-1))}}(\boldsymbol{a}_k + \boldsymbol{a}_l): k<l\), and \(l = 1,2,\cdots,n+1.\)
Finally, using (\ref{Eq: Spherical Radial Rule})-(\ref{Eq: Spherical Simplex}), the integral \(I(g)\) in (\ref{Eq: I of g}) can be approximated using the stochastic spherical-radial rule as expressed in (\ref{Eq:I(g)16})-(\ref{Eq:I(g)17}), where \(\varpi_0 = 1 - {n(\rho_1^2 + \rho_2^2 - (n+2))}/{(\rho_1^2\rho_2^2)}\) and \(\bar{g}(\boldsymbol{x}) = \frac{1}{2}(g(\boldsymbol{x}) + g(\boldsymbol{-x}))\).

\begin{table*}
\centering
\begin{minipage}{0.9\textwidth}
\begin{align}\nonumber
I(g) &\approx \varpi_{s,1}\sum_{j=1}^{n+1}\bigg[\varpi_{r,0}g(\boldsymbol{0}) + \varpi_{r,1}\frac{g(-\rho_1\mathcal{Q}\boldsymbol{a}_j) + g(\rho_1\mathcal{Q}\boldsymbol{a}_j)}{2}  + \varpi_{r,2}\frac{g(-\rho_2\mathcal{Q}\boldsymbol{a}_j) + g(\rho_2\mathcal{Q}\boldsymbol{a}_j)}{2}\bigg] \\ & \label{Eq:I(g)16} + \varpi_{s,2}\sum_{j=1}^{{n(n+1)}/{2}}\bigg[\varpi_{r,0}g(\boldsymbol{0}) + \varpi_{r,1}\frac{g(-\rho_1\mathcal{Q}\boldsymbol{b}_j) + g(\rho_1\mathcal{Q}\boldsymbol{b}_j)}{2} + \varpi_{r,2}\frac{g(-\rho_2\mathcal{Q}\boldsymbol{b}_j) + g(\rho_2\mathcal{Q}\boldsymbol{b}_j)}{2}\bigg]\\ \label{Eq:I(g)17}
& \approx \varpi_0g(\boldsymbol{0}) + \varpi_{s,1}\sum_{j=1}^{n+1}\bigg[\varpi_{r,1}\bar{g}(\rho_1\mathcal{Q}\boldsymbol{a}_j) + \varpi_{r,2}\bar{g}(\rho_2\mathcal{Q}\boldsymbol{a}_j)\bigg] + \varpi_{s,2}\sum_{j=1}^{{n(n+1)}/{2}}\bigg[\varpi_{r,1}\bar{g}(\rho_1\mathcal{Q}\boldsymbol{b}_j) + \varpi_{r,2}\bar{g}(\rho_2\mathcal{Q}\boldsymbol{b}_j)\bigg]
\end{align}
\hrule
\end{minipage}
\end{table*}

\smallskip
\textit{Remark 4}: To achieve global convergence, the stochastic integration is evaluated \(N_{m}\) times and averaged. In each evaluation, independent realizations of random entities \(\rho_1\), \(\rho_2\) and \(\mathcal{Q}\) are considered. From (\ref{Eq:I(g)17}), we note that each iteration operates for \(n^2 + 3n +3\) points.  Hence, the total number of function evaluations required is \(N_{m}( n^2 + 3n + 3)\).

\section{Stochastic Integration Filtering}\label{SectionSIF}
Here, we describe the procedure to recursively estimate \(\hat{\boldsymbol{x}}_{k|k}\) using the stochastic integration rule described in Section \ref{subsec: Fifth-degree rule}. The filter is initialized with \(\hat{\boldsymbol{x}}_{0|0} = \EE[\boldsymbol{x}_0]\) and \(P_{0|0}  = E[(\boldsymbol{x}_0 - \hat{\boldsymbol{x}}_{0|0})(\boldsymbol{x}_0 - \hat{\boldsymbol{x}}_{0|0})^T]\). The filtering procedure is carried out by repeating the following steps for each instance \(k\).

For the \textit{state prediction step}, we set  \(\boldsymbol{\mu} = \hat{\boldsymbol{x}}_{k-1|k-1}\), \(\Sigma = P^{xx}_{k-1|k-1}\) and generate independent realizations of \(\rho^l_1\), \(\rho^l_2\) and \(\mathcal{Q}^l\) for \(l= 1,2,\cdots,N_m\). Then, for each \(l\), we generate the following set of sigma-points for \(j=1,2\):
\begin{subequations}
\label{eq: Sigma points}
\begin{alignat}{1}
\boldsymbol{X}^l_{i,a,\rho_j} &= \boldsymbol{\mu} + \sqrt{\Sigma}\rho^l_j\mathcal{Q}^l\boldsymbol{a}_i \qquad 0<i \leq n+1,  \\
\boldsymbol{X}^l_{i,b,\rho_j} &= \boldsymbol{\mu} + \sqrt{\Sigma}\rho^l_j\mathcal{Q}^l\boldsymbol{b}_i \qquad 0<i \leq n(n+1)/2.
\end{alignat}
\end{subequations}
Let \(f_1(\boldsymbol{x}) = f(\boldsymbol{x})\), \(f_2(\boldsymbol{x}) = f(\boldsymbol{x})f(\boldsymbol{x})^T\), and \(\bar{f}_i(\boldsymbol{x}) = \frac{1}{2}({f_i(\boldsymbol{x}) + f_i(-\boldsymbol{x})})\), for \(i=1,2\). Then, using (\ref{Eq:I(g)17}), the integrals in (\ref{Eq: x hat k given k}) and (\ref{eq: P k given k-1}) are approximated as
\begin{subequations}
\begin{alignat}{1}\nonumber
\hat{\boldsymbol{x}}_{k|k-1} &= \frac{1}{N_m}\sum\limits_{l=1}^{N_m}\bigg[f(\boldsymbol{\mu})\varpi^l_0 + \varpi_{s,1}\sum_{i=1}^{n+1}\sum_{j=1}^{2}\bar{f}_1(\boldsymbol{X}^l_{i,a,\rho_j})\varpi^l_{r,j} \\ & + \varpi_{s,2}\sum_{i=1}^{n(n+1)/2}\sum_{j=1}^{2}\bar{f}_1(\boldsymbol{X}^l_{i,b,\rho_j})\varpi^l_{r,j}\bigg], \\ \nonumber P^{xx}_{k|k-1} &= \frac{1}{N_m}\sum\limits_{l=1}^{N_m}\bigg[ f(\boldsymbol{\mu})f(\boldsymbol{\mu})^T\varpi^l_0 + \varpi_{s,1}\sum_{i=1}^{n+1}\sum_{j=1}^{2}\bar{f}_2(\boldsymbol{X}^l_{i,a,\rho_j})\varpi^l_{r,j} \\ & + \varpi_{s,2}\sum_{i=1}^{n(n+1)/2}\sum_{j=1}^{2}\bar{f}_2(\boldsymbol{X}^l_{i,b,\rho_j})\varpi^l_{r,j}\bigg] \!\!+ \!Q_k \! - \!\hat{\boldsymbol{x}}_{k|k-1}\hat{\boldsymbol{x}}_{k|k-1}^T.
\end{alignat}
\end{subequations}
For the \textit{observation prediction step}, we set  \(\boldsymbol{\mu} = \hat{\boldsymbol{x}}_{k|k-1}\), \(\Sigma = P^{xx}_{k|k-1}\) and generate a new set of sigma-points using (\ref{eq: Sigma points}). Let  \(h_1(\boldsymbol{x}) = h(\boldsymbol{x})\), \(h_2(\boldsymbol{x}) = \boldsymbol{x}h(\boldsymbol{x})^T\), \(h_3(\boldsymbol{x}) = h(\boldsymbol{x})h(\boldsymbol{x})^T\) and \(\bar{h}_i(\boldsymbol{x}) = \frac{1}{2}(h_i(\boldsymbol{x}) + h_i(-\boldsymbol{x}))\), \(i=1,2,3\). Now  using (\ref{Eq:I(g)17}), the integrals in (\ref{eq: y hat k given k-1}), (\ref{eq: P xy k given k-1}) and (\ref{Eq: P yy k given k minus 1}) are approximated as
\begin{subequations}
\begin{alignat}{1}
& \nonumber\hat{\boldsymbol{y}}_{k|k-1} = \frac{1}{N_m}\sum\limits_{l=1}^{N_m}\bigg[h(\boldsymbol{\mu})\varpi^l_0 + \varpi_{s,1}\sum_{i=1}^{n+1}\sum_{j=1}^{2}\bar{h}_1(\boldsymbol{X}^l_{i,a,\rho_j})\varpi^l_{r,j} \\ &\qquad + \varpi_{s,2}\sum_{i=1}^{n(n+1)/2}\sum_{j=1}^{2}\bar{h}_1(\boldsymbol{X}^l_{i,b,\rho_j})\varpi^l_{r,j}\bigg], \\
&\nonumber \hat{P}^{xy}_{k|k-1} = \frac{1}{N_m}\sum\limits_{l=1}^{N_m}\bigg[ \boldsymbol{\mu} h(\boldsymbol{\mu})^T\varpi^l_0 + \varpi_{s,1}\sum_{i=1}^{n+1}\sum_{j=1}^{2}\bar{h}_2(\boldsymbol{X}^l_{i,a,\rho_j})^T\varpi^l_{r,j} \\
&\qquad + \varpi_{s,2}\sum_{i=1}^{n(n+1)/2}\sum_{j=1}^{2}\bar{h}_2(\boldsymbol{X}^l_{i,b,\rho_j})^T\varpi^l_{r,j}\bigg] - \hat{\boldsymbol{x}}_{k|k-1}\hat{\boldsymbol{y}}_{k|k-1}^T, \\
&\nonumber \hat{P}^{yy}_{k|k-1} = \frac{1}{N_m}\sum\limits_{l=1}^{N_m}\bigg[ h(\boldsymbol{\mu})h(\boldsymbol{\mu})^T\varpi^l_0 + \varpi_{s,1}\sum_{i=1}^{n+1}\sum_{j=1}^{2}\bar{h}_3(\boldsymbol{X}^l_{i,a,\rho_j})\varpi^l_{r,j} \\
&\qquad + \varpi_{s,2}\sum_{i=1}^{n(n+1)/2}\sum_{j=1}^{2}\bar{h}_3(\boldsymbol{X}^l_{i,b,\rho_j})\varpi^l_{r,j}\bigg] - \hat{\boldsymbol{y}}_{k|k-1}\hat{\boldsymbol{y}}_{k|k-1}^T +R_k.
\end{alignat}
\end{subequations}
Finally, the correction step follows (\ref{Eq: x k given k})-(\ref{Eq: P xx k given k}).

\section{Simulation results}
\label{sec: Simulations}
In this Section, we compare the performance of the proposed SIF with the third-degree SIF, third- and fifth-degree CKF, and fifth-degree quasi-stochastic filter \cite{zhang2014quasi}. The first example considers approximating a nonlinear integral; whereas, the second example considers a filtering scenario.

\subsection{Approximating a Nonlinear Integral}
Let \(\boldsymbol{x} = [x_1,x_2,\cdots,x_n]^T\) be a random vector consisting of zero-mean independent Gaussian variables, i.e., \(\boldsymbol{x} \sim \mathcal{N}_{\boldsymbol{x}}(\boldsymbol{0},I)\). We consider a Gaussian weighted integral of the form \(I(g) = \int g(\boldsymbol{x})\mathcal{N}_{\boldsymbol{x}}(\boldsymbol{0},I)\textrm{d}\boldsymbol{x}\), where \(g(\boldsymbol{x}) = \sum_{i=1}^{n}x^i_i\). The true value of the integral is \(I_T = \sum_{p=1}^{n}(p-1)!!I_{i}(p)\), where \(!!\) denotes double factorial and \(I_{i}(p)\) is an indicator function that returns \(0\) if \(p\) is odd and \(1\) if \(p\) is even.

For \(n=6\) and consequently \(I_T = 19\), the relative error, defined as \(R_e = |I_T - I_{A}|/I_T\), for various approximation methods is tabulated in Table \ref{tab:table1}, where \(I_{A}\) is the approximate value obtained by the various integration rules. We provide the maximum and the average error values for the stochastic methods obtained after \(1000\) runs. The deterministic methods, i.e., the third- and fifth-degree CKF, have the same value of the maximum and average error, hence only average values are shown. The value of \(N_{m}\) is adjusted such that all stochastic integration methods utilize approximately the same number of points. We observe that, for the given scenario, both third- and fifth-degree CKF give unreliable approximations and have very large values of relative errors. The stochastic methods, on the other hand, provide superior average performances and the proposed fifth-degree SIF outperforms all other filters. Furthermore, the third-degree SIF is found to have a very large value of maximum relative error, and hence, may occasionally give large errors in filtering applications. Moreover, we employed Monte-Carlo integration, where \(I(g)\) is approximated using the average of \(600\) random realizations of \(g(\boldsymbol{x})\); note that it performed far inferior to proposed scheme.
\begin{table}[!ht]
	\centering
	\caption{Relative errors of addressed integral rules}
	\label{tab:table1}
	\begin{tabular}{ccccc}
		\hline
		Rule  & \(R_{e,\textrm{max}}\) \% & \(R_{e,\textrm{mean}}\) \% & \(N_{m}\) & Points\\
		\hline
		Third-degree CKF & --- & 104.0521 & ---& 12\\
		Fifth-degree CKF & --- & 57.89 & ---& 56\\
		Third-degree SIF &  83.11 & 13.92 & 50 & 600\\
		Fifth-degree SIF &  24.98 & 6.43 & 10 & 570\\
		Fifth-degree QSIF &  23.68 & 15.89 & 10 & 560\\
		Monte-Carlo Integration & 99.25 & 18.33 & --- & 600\\
		 		
	\end{tabular}
\end{table}

\subsection{Nonlinear Filtering}
\begin{figure}[!htbp]\centering
	\includegraphics[scale=0.55, bb=31 108 553 688, clip]{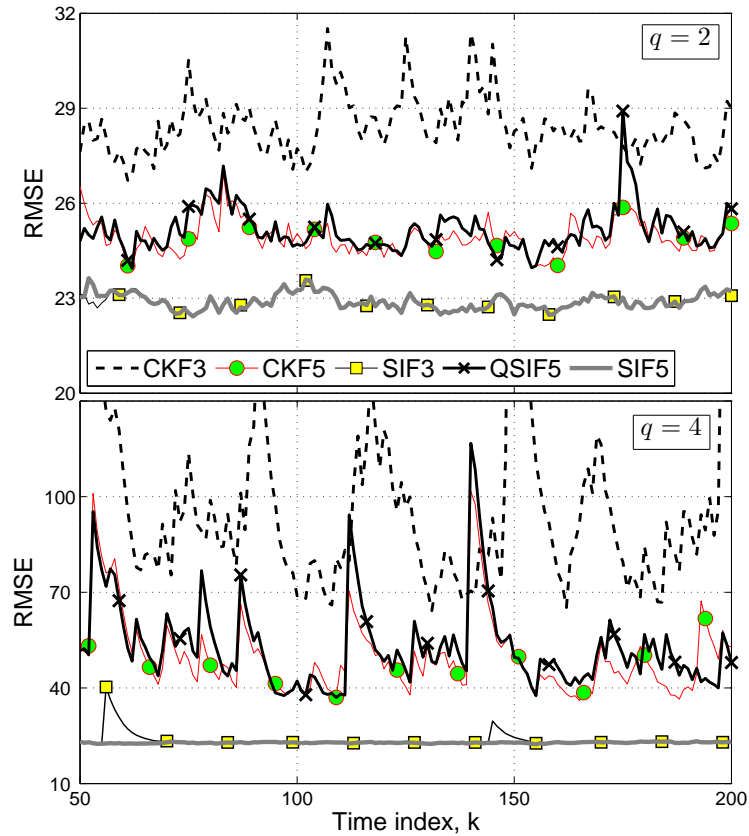}
	\caption{Comparison of RMSE of the proposed fifth-degree SIF (SIF5) with third-degree SIF (SIF3), third-degree CKF (CKF3), fifth-degree CKF (CKF5), and fifth-degree QSIF (QSIF5) for \(q=2\) (above), and \(q=4\) (below).}\label{fig:RMSE_q_2}
\end{figure}

We consider the following state-space model \cite{zhang2014seventh}
\begin{subequations}
\label{eq: State-space model simulation example}
\begin{alignat}{1}
\boldsymbol{x}_k &= 0.9\boldsymbol{x}_{k-1} + \boldsymbol{w}_k, \\
y_k &= z_k^q + v_k,
\end{alignat}
\end{subequations}
where \(z_k = (1 + \boldsymbol{x}_k^T\boldsymbol{x}_k)^2\), \(\boldsymbol{w}_k \sim \mathcal{N}(\boldsymbol{0},Q)\) with \(Q = 100I_n\) and \(n=10\), and \(v_k \sim \mathcal{N}(0,R)\) with \(R = 10\). The filter is initialized with \(\hat{\boldsymbol{x}}_{0|0} = \EE[\boldsymbol{x}_0]\), where \(\boldsymbol{x}_0 \sim \mathcal{N}(\boldsymbol{1}_{n \times 1},P^{xx}_{0|0})\) and \(P^{xx}_{0|0} = 10I_n\). The parameter \(q\) can be tuned to adjust the degree of nonlinearity in the state-space model. We have carried out the simulation experiments for various values of \(q\). We compare the performance of various filters using root-mean-square-error (RMSE) as the performance metric, the RMSE is obtained using the following relation:\begin{small}
\begin{equation}
\label{Eq: RMSE}
\textrm{RMSE}_k = \sqrt{\frac{1}{N_{\textrm{MC}}}\sum_{m=1}^{N_{\textrm{MC}}}||\hat{\boldsymbol{x}}_{k|k,m} - \boldsymbol{x}_k||^2_2},
\end{equation}\end{small}
 where \(N_{\textrm{MC}} = 500\). The parameter \(N_{m}\) is set equal to \(10\) for both fifth-degree SIF and QSIF; while, it is \(50\) for the third-degree SIF. In Fig.~\ref{fig:RMSE_q_2} (above) for \(q=2\), we observe that the fifth-degree CKF and QSIF have similar performances, and they perform better than the third-degree CKF; the third- and fifth-degree SIFs, however, outperform the fifth-degree CKF and QSIF. In Fig.~\ref{fig:RMSE_q_2} (below) for \(q=4\), we observe that, all filters exhibit large peaks in their respective RMSE values, but that of proposed fifth-degree SIF remains stable and smaller.
\section{Conclusion}
\label{sec: Conclusions}
In this letter, we discussed the utilization of higher-degree spherical-radial stochastic integration rules for nonlinear Bayesian filtering. We specifically developed a fifth-degree stochastic integration filter (SIF). The performance of the proposed filter was compared with the third- and fifth-degree cubature Kalman filter, the third-degree SIF, and the fifth-degree quasi-SIF for a nonlinear filtering scenario. It was observed that the proposed fifth-degree SIF can perform better than existing ones.

\vspace{5mm}

\bibliographystyle{IEEEtran}
\bibliography{ref}

\end{document}